\documentclass[12pt]{article}

\usepackage{braket}
\usepackage{mathrsfs}
 \usepackage{extpfeil}
\usepackage[nosort]{cite}
 \usepackage{graphics}
\usepackage{epstopdf}

\numberwithin{equation}{section}
\begin{document}

\begin{center}
{\LARGE{\bf{From the Weyl quantization of a particle on the circle to number-phase Wigner functions }}}
\end{center}

\bigskip\bigskip

\begin{center}
Maciej Przanowski\footnote{E-mail address:  maciej.przanowski@p.lodz.pl}, Przemys\l aw Brzykcy\footnote{E-mail address:  800289@edu.p.lodz.pl} and Jaromir Tosiek\footnote{E-mail address:  jaromir.tosiek@p.lodz.pl}
\end{center}

\begin{center}

{\sl  Institute of Physics, Technical University of  \L\'{o}d\'{z},\\ W\'{o}lcza\'{n}ska 219, 90-924 \L\'{o}d\'{z}, Poland.}\\
\medskip

\end{center}

\vskip 1.5cm
\centerline{\today}
\vskip 1.5cm

\begin{abstract}
A generalized Weyl quantization formalism for a particle on the circle investigated in \cite{1} is developed. A Wigner function for the state $\hat{\varrho}$  and the kernel $\mathcal{K}$ for a particle on the circle is defined and its properties are analyzed. Then it is shown how this Wigner function can be easily modified  to give the number-phase Wigner function in quantum optics. Some examples of such number-phase Wigner function are considered.
\end{abstract}



 \section{Introduction} \label{sec1} 

The present paper can be regarded as a continuation of our previous work \cite{1}. In fact, it has been motivated by a question raised by the referee of \cite{1}. Namely, the referee pointed out that the Weyl quantization formalism developed in \cite{1} should be closely related to the Wigner function and the Wigner representation of quantum phase investigated previously by the others \cite{2,3,4}. So, here we follow this suggestion and we intend to display, how one can define the Wigner function which depends on the number and the phase. We arrive at this goal by employing the generalized Weyl quantization formalism for a particle on the circle given in \cite{1}. As will be shown the answer to this question can be easily found and, moreover, it appears to be fairly natural within the generalized Weyl quantization machinery. 
It is well known that the Wigner function for a system of particles in $\mathbb{R}^3$ is a real function on  the corresponding classical phase space so it depends on the Cartesian coordinates of the particles and on the respective canonically conjugate momenta \cite{5,6,7}. The Wigner function is uniquely  defined by the density operator of the system and, conversely, the density operator is uniquely determined by the given Wigner function. 
Then, the expectation value of any quantum observable in a given quantum state can be found by integrating the product of the corresponding classical observable and the respective Wigner function. This procedure resembles very much the one well known in statistical physics. 
The only important difference consists in the fact that, in general, the Wigner function is not pointwise non-negative.  
Consequently, in general, it does not represent a probability distribution but, as is usually said, the Wigner function is a \textit{quasiprobability distribution}. 
Nevertheless, the marginal distributions of any Wigner function give the probability distributions for coordinates and momenta, respectively. 
The natural question arises if one can define the analogous Wigner functions for the constraint quantum systems (e.g. a particle on the circle) or for the quantum systems described by the finite-dimensional Hilbert spaces. 
This question has raised a great deal of interest and many authors have analysed various examples of the Wigner functions for such quantum systems \cite{8,9,10,11,12,13,14,15,16,17,18} (in particular see \cite{17} and the references therein).
Another interesting question, which turns out to be closely related to the previous one, concerns the definition of the Wigner function in quantum optics as a function of the photon number and the phase. As is known, the photon number and the phase can be considered as the canonically conjugated quantities and therefore, one attempts to find the Wigner function that depends on these quantities i.e. the so called \textit{number-phase Wigner function}. 
This problem was explored by numerous authors. For example, J. A. Vaccaro and D. T. Pegg \cite{19} defined a  number-phase Wigner function employing the results of W. K. Wootters on the discrete Wigner functions in finite-dimensional spaces \cite{12}. Such an approach is undoubtedly based on the celebrated Pegg-Burnett formalism in the theory of quantum phase \cite{21,22,23}. 
A. Luk\v s with V. Pe\v rinov\'a \cite{2}, and M. R. Hush \textit{et al} \cite{4} defined a number-phase Wigner function which was extended to rather unphysical half integer values of the number of photons. 
In his distinguished work J. A. Vaccaro \cite{3} has found a number-phase Wigner function assuming at the very beginning that this function should  satisfy, besides the usual properties inflicted on Wigner function \cite{7}, some additional property leading to the interference fringes for the Wigner functions of the Schr\"odinger cat states. 

The aim  of our paper is to show that the number-phase Wigner function can be easily defined with the use of the generalized Weyl quantization formalism on the cylindrical phase space $S^1\times \mathbb{R}^1$ under the observation that Hilbert space $L^2(S^1)$ can be considered as an enlarged Hilbert space of the Fock space $\mathcal{H}_F$.  
The paper is organized as follows. In Section \ref{sec2} we develop the generalized Weyl quantization formalism for the cylindrical phase space $S^1\times\mathbb{R}^1$. The generalized Stratonovich-Weyl quantizer for an arbitrary kernel $\mathcal{K}$ is introduced and its properties are studied. The generalized Weyl symbol of operator is defined and its basic properties are investigated.  
In Section \ref{sec3} the generalized Moyal star product in $S^1\times\mathbb{R}^1$ is considered.
The definition of the Wigner function (for an arbitrary kernel $\mathcal{K}$) in $S^1\times\mathbb{R}^1$ is given in Section \ref{sec4}.
Main properties of the Wigner functions are also analysed. 
The eigenvalue equations for the Wigner functions are presented in Section \ref{sec5}. 
In Section \ref{sec6} the Wigner function is specified to the case when the kernel $\mathcal{K}=\mathcal{K}_S$ leads to the symmetric ordering of operators under the Weyl quantization prescription.  
The number-phase Wigner function is defined in Section \ref{sec7}.
We show there that having defined the Wigner function in $S^1\times\mathbb{R}^1$ for the kernel $\mathcal{K}=\mathcal{K}_S$
and employing the results of our previous work on quantum phase \cite{1} one can quickly define a number-phase Wigner function in quantum optics. The properties of this Wigner functions are also studied in Section \ref{sec7}. 
Section \ref{sec8} is devoted to some explicit examples of the number-phase Wigner functions.
We  consider the Fock states, the coherent states, the squeezed states, the black body radiation and the `Fock cat' states.
Finally, concluding remarks in Section \ref{sec9} end the paper.

\section{The generalized  Weyl quantization on the cylinder and the generalized Weyl symbol}\label{sec2}

Consider a particle on the circle $S^1$. Let $\Theta \in [-\pi,\pi)$ denote the angle coordinate of the particle [\textit{Important remark:} In our paper the interval $[-\pi,\pi)$ is identified with the circle $S^1$ ]  and $L\in \mathbb{R}^1$
be the angular momentum of this particle. The corresponding phase space is the cylinder $S^1 \times \mathbb{R}^1$.  
 Given a function  $f=f(\Theta , L)$ on  $S^1 \times \mathbb{R}^1$ one assigns to it  an operator $W[\mathcal{K}](f)$
in the Hilbert space $L^2(S^1)$ according to the rule \cite{24,1}
\begin{equation}
\begin{split}
&f =f(\Theta,L)\mapsto W[\mathcal{K}] (f)\\
W[\mathcal{K}] (f) :=& 
\sum_{l=-\infty}^{\infty} \int_{-\pi} ^{\pi}  \mathcal{K}(\sigma, l)  \Big\{ \sum_{n=-\infty}^{\infty} \int_{-\pi}^{\pi}   f(\Theta , n\hslash) \exp{\big\{ -i(\sigma n + l \Theta) \big \}} \frac{\mathrm{d} \Theta}{2\pi} \Big\} \hat{U}(\sigma,l) \frac{\mathrm{d} \sigma}{2\pi} \\
=& \sum_{n=-\infty}^{\infty} \int_{-\pi}^{\pi} f(\Theta, n \hslash ) \hat{\Omega}[\mathcal{K} ] (\Theta, n) \frac{\mathrm{d} \Theta}{2\pi},
\end{split}\label{2.1}
\end{equation}
where the  function $\mathcal{K} =\mathcal{K}(\sigma, l)$, $\sigma\in [-\pi,\pi)$, $l\in\mathbb{Z}$, called the \textit{kernel},    is   smooth with respect to $\sigma$    and $\hat{U}(\sigma,l)$ is the unitary operator on the Hilbert space $L^2(S^1)$
\begin{equation}
\begin{split}
\hat{U} (\sigma,l) &= \exp{\Big\{ i \Big( \frac{\sigma}{\hslash} \hat{L} + l\hat{\Theta} \Big)\Big\}}\\
&= \exp{\Big\{ \frac{i}{2} \sigma l  \Big\}} \exp{\Big\{ il\hat{\Theta}  \Big\}} \exp{\Big\{ \frac{i}{\hslash} \sigma \hat{L} \Big\}}\\
 &= \exp{\Big\{ -\frac{i}{2} \sigma l  \Big\}}  \exp{\Big\{ \frac{i}{\hslash} \sigma \hat{L} \Big\}}\exp{\Big\{ il\hat{\Theta}  \Big\}}\\
&= \sum_{k=-\infty}^{\infty} \exp{ \Big\{ i\sigma \Big( k+\frac{l}{2}\Big) \Big\}  } \ket{k+l}\bra{k},\quad \sigma\in[-\pi,\pi).\label{2.2}
\end{split}
\end{equation}
Moreover,
\begin{equation}
\begin{split}
\hat{\Omega}[\mathcal{K}](\Theta, n) := \sum_{l=-\infty}^{\infty} \int_{-\pi}^{\pi} \mathcal{K}(\sigma,l)\exp{\Big\{ -i \big( \sigma n + l \Theta\big) \Big\} } \hat{U}(\sigma,l) \frac{\mathrm{d} \sigma}{2\pi}\label{2.3}
\end{split}
\end{equation}
is the \textit{generalized Stratonovich-Weyl (GSW) quantizer for the kernel $\mathcal{K}$}.
Recall that $\hat{U}(\sigma, l)$ has the following properties \cite{16,16a} 
\begin{subequations}
\begin{eqnarray}
\mathrm{Tr} \Big\{ \hat{U}(\sigma,l)\Big\} &=& 2\pi \delta_{l0} \delta^{(S)}(\sigma),\label{2.4a}\\
\mathrm{Tr}\Big\{\hat{U}^{\dagger}(\sigma,l)\hat{U}(\sigma',l') \Big\} &=& 2\pi \delta_{ll'}\delta^{(S)}(\sigma-\sigma'),\label{2.4b}
\end{eqnarray}
\end{subequations}
where $\delta^{(S)}(\sigma)$ stands for the Dirac delta on the circle given by
\begin{equation}
\delta^{(S)} (\sigma) =\frac{1}{2\pi} \sum_{l=-\infty}^{\infty} \exp{\big\{ il\sigma\big\}}.
\end{equation}
As can be easily shown (see for example \cite{6,1} ) the natural assumptions about quantization impose some restrictions on the kernel $\mathcal{K}$. Namely 
\begin{subequations}
\begin{enumerate}
\item[(i)] $W[\mathcal{K}](f) = f(\hat{\Theta})$ for an arbitrary function $f$ depending only on $\Theta$, $f=f(\Theta)$ iff 
\begin{equation}
\forall_{l\in\mathbb{Z} }\quad \mathcal{K}(0,l)=1,\label{2.6a}
 \end{equation}
\item[(ii)] $W[\mathcal{K}](f) = f(\hat{L})$ for an arbitrary function $f$ depending only on $L$, $f=f(L)$ iff 
\begin{equation}
\forall_{\sigma \in [-\pi,\pi) }\quad \mathcal{K}(\sigma,0)=1,\label{2.6b}
 \end{equation}
\item[(iii)] the operator $W[\mathcal{K}]$ is symmetric for any real function $f=f(\Theta,L)$ iff
\begin{equation}
\forall_{\sigma \in [-\pi,\pi),\,\,l\in\mathbb{Z} }\quad \mathcal{K}^*(\sigma,l)=\mathcal{K}(-\sigma,-l)\label{2.6c}
 \end{equation}
(where the star `$*$' stands for the complex conjugation). 
\end{enumerate} 
\end{subequations}
Then the GSW quantizer (\ref{2.3}) has the following important properties:
\begin{subequations}
\begin{enumerate}
\item[(a)] If (\ref{2.6a}) or (\ref{2.6b}) hold true then
\begin{equation}
\label{2.7a}
\mathrm{Tr} \left\{ \hat{\Omega} [\mathcal{K}] (\Theta,n) \right\} =1.
\end{equation}
\item[(b)] The condition (\ref{2.6c}) yields
\begin{equation}
\hat{\Omega}^{\dagger} [\mathcal{K}] =\hat{\Omega} [\mathcal{K}] .
\label{2.7b}
\end{equation}
\item[(c)]
\begin{equation}
\label{2.7c}
\begin{split}
&\mathrm{Tr} \left\{ \hat{\Omega} [\mathcal{K}] (\Theta,n)\hat{\Omega} [\mathcal{K}] (\Theta',n')\right\}  =\\
&= \frac{1}{2\pi} \sum_{l=-\infty}^{\infty}  \int_{-\pi}^{\pi} \mathcal{K}(\sigma,l)\mathcal{K}(-\sigma,-l)\exp{\left\{ i [\sigma(n-n') + l(\Theta-\Theta')] \right\} \mathrm{d} } \sigma.
\end{split}
\end{equation}
\item[(d)] If (\ref{2.6a}) holds true then
\begin{equation}
\label{2.7d}
\frac{1}{2\pi} \sum_{n=-\infty}^{\infty}\hat{\Omega} [\mathcal{K}] (\Theta,n) = \ket{\Theta}\bra{\Theta}.
\end{equation}
\item[(e)] If (\ref{2.6b}) holds true then
\begin{equation}
\label{2.7e}
\frac{1}{2\pi} \int_{-\pi}^{\pi}\hat{\Omega} [\mathcal{K}] (\Theta,n)\mathrm{d} \Theta = \ket{n}\bra{n}.
\end{equation}
\end{enumerate}
\end{subequations}
Carrying out the procedure involving changing $\sigma \to \sigma'$ and $l\to l'$ in Eq. (\ref{2.1}), then multiplying both sides of this equation by $\hat{U}^{\dagger}(\sigma,l)$, taking the trace and employing the formula (\ref{2.4b}) one arrives at 
\begin{equation}
\begin{split}
\mathrm{Tr}  \left\{ \hat{U}^{\dagger}(\sigma,l)  W[\mathcal{K}] (f)\right\}= \mathcal{K}(\sigma,l) \sum_{n=-\infty} ^ {\infty} \int_{-\pi} ^ {\pi} f(\Theta,n\hslash) \exp{\left\{ -i(\sigma n + l\Theta ) \right\}} \frac{\mathrm{d} \Theta}{2\pi}.
\label{2.8}
\end{split}
\end{equation}
Assume that 
\begin{equation}
\label{2.9}
\forall_{\sigma \in [-\pi,\pi),\,\, l\in \mathbb{Z}}\quad \mathcal{K}(\sigma,l) \neq 0,
\end{equation}
then we can extract $f(\Theta,n\hslash)$ from (\ref{2.8}). Namely
\begin{equation}
\label{2.10}
f(\Theta,n\hslash) = \sum_{l=-\infty}^{\infty} \int_{-\pi}^{\pi} 
\left( \mathcal{K}(\sigma,l) \right)^{-1} \mathrm{Tr} \left\{ \hat{U}^{\dagger}(\sigma,l) W[\mathcal{K}](f)  \right\} \exp{\left\{ i(\sigma n + l\Theta) \right\} } 
\frac{\mathrm{d} \sigma}{2\pi}.
\end{equation}
This last result leads to the following conclusion. 
Given an operator $\hat{f}$ in the Hilbert space $L^2(S^1)$ one can define a function on the quantized cylindrical phase space $S^1\times \mathbb{Z}$ according to the rule 
\begin{equation}
\begin{split}
\hat{f} \mapsto W^{-1} [\mathcal{K}] (\hat{f} ) (\Theta,n\hslash) :  = \sum_{l=-\infty} ^{\infty} \int_{-\pi}^{\pi} \left( \mathcal{K}(\sigma,l) \right)^{-1} \mathrm{Tr} 
\left\{ \hat{U}^{\dagger}(\sigma,l) \hat{f} \right\}
 \exp{\left\{ i(\sigma n+l\Theta\right\}}
 \frac{\mathrm{d} \sigma}{2\pi}.
\end{split}
\label{2.11}
\end{equation}
This function we call the  \textit{generalized Weyl symbol of $\hat{f}$ for the kernel $\mathcal{K}$}.
The following relations are obvious
\begin{subequations}
\begin{eqnarray}
W^{-1}[\mathcal{K}] \left( W[\mathcal{K} ] (f) \right) &=& f, \label{2.12a}\\
 W[\mathcal{K}] \left( W^{-1}[\mathcal{K} ] (\hat{f}) \right) &=& \hat{f}. \label{2.12b}
\end{eqnarray}
\end{subequations}
In particular, for $\mathcal{K}=1$ the function $W^{-1}[1](\hat{f})(\Theta,n\hslash)$ is simply the \textit{Weyl symbol of $\hat{f}$} and from (\ref{2.11}) with (\ref{2.3}) on has 
\begin{equation}
W^{-1}[1](\hat{f}) (\Theta,n\hslash) = \mathrm{Tr} \left\{ \hat{\Omega}[1](\Theta,n) \hat{f} \right\}. \label{2.13}
\end{equation}
To proceed further and to simplify the formulas we introduce  an operator $\hat{K}$, defined by
\begin{equation}
\hat{K} F(\theta,n) := \sum_{n'=-\infty} ^{\infty} \int_{-\pi}^{\pi} K(\Theta,n;\Theta',n') F(\Theta',n') \mathrm{d}\Theta',
\label{2.14}
\end{equation}
where
\begin{equation}
K(\Theta,n;\Theta',n'):= \frac{1}{4\pi^2}  \sum_{l=-\infty} ^{\infty} \int_{-\pi}^{\pi} 
\mathcal{K}(\sigma,l)
\exp{\left\{ -i  \left[   \sigma(n-n') + l(\Theta-\Theta') \right] \right\}}
\mathrm{d} \sigma
\label{2.15}
\end{equation}
(note that $\hat{K}$ is the counterpart of the operator $\alpha\left( -\hslash \frac{\partial^2}{\partial p  \partial q} \right)$ employed in \cite{25} for the case of the phase space $\mathbb{R}^1\times \mathbb{R}^1 $). 
If the condition (\ref{2.6c}) is fulfilled then 
\begin{equation}
K^*(\Theta, n; \Theta',n')=K(\Theta', n'; \Theta,n)
\label{2.16}
\end{equation}
and it means that the operator $\hat{K}$ is Hermitian 
\begin{equation}
\hat{K}^{\dagger} = \hat{K}.
\label{2.17}
\end{equation}
In general, independently whether (\ref{2.17}) is satisfied or not, Eq. (\ref{2.3}) can be rewritten in a following concise form 
\begin{equation}
\hat{\Omega}[\mathcal{K}] = \hat{K}\hat{\Omega}[1].
\label{2.18}
\end{equation}
Then the generalized Weyl symbol (\ref{2.11}) takes the form
\begin{equation}
\begin{split}
&W^{-1}[\mathcal{K}] (\hat{f} ) (\theta,n\hslash) = \hat{K}^{-1}\cdot \left( \hat{ {K}}^T\right)^{-1}
\mathrm{Tr} \left\{  \hat{\Omega} [\mathcal{K} ](\Theta,n)   \hat{f}\right\}\\
&\stackrel{\mathrm{by}\, (\ref{2.18})}{=}
 \left( \hat{ {K}}^T\right)^{-1} \mathrm{Tr} \left\{  \hat{\Omega} [1] (\Theta,n)   \hat{f}\right\}\\
 &\stackrel{\mathrm{by}\, (\ref{2.13})}{=}
 \left( \hat{ {K}}^T\right)^{-1} W^{-1}[1](\hat{f})(\Theta,n\hslash),
\end{split}
\label{2.19}
\end{equation}
where $\hat{K}^T$ is the operator transposed to $\hat{K}$ i.e. 
\begin{equation}
\hat{K}^T F(\theta,n) := \sum_{n'=-\infty} ^{\infty} \int_{-\pi}^{\pi} K(\Theta',n';\Theta,n) F(\Theta',n') \mathrm{d}\Theta'
\label{2.20}
\end{equation}
(compare with Eq. ($4.23$) of \cite{25}).
Let $\hat{f}$ be an operator in $L^2(S^1)$ which arises from some function $f(\Theta,n\hslash)$ on the quantized cylindrical phase space $S^1\times \mathbb{Z}$ by the generalized Weyl quantization prescription (\ref{2.1}) and let $\hat{g}$ be any operator in $L^2(S^1)$. Then from (\ref{2.1}) one quickly gets 
\begin{equation}
\begin{split}
\mathrm{Tr} \left\{ \hat{f} \hat{g} \right\} &= \mathrm{Tr}
 \left\{ 
 \left[ 
 	\sum_{n=-\infty} ^{\infty} \int_{-\pi} ^{\pi} 
 	f(\Theta,n\hslash) \hat{\Omega}[\mathcal{K}] 	(\theta,n )	 
 	 \frac{\mathrm{d}\Theta}{2\pi}  
 \right]   \hat{g}  
 \right\} \\
 &=
 	\sum_{n=-\infty} ^{\infty} \int_{-\pi} ^{\pi} 
 	f(\Theta,n\hslash)\frac{1}{2\pi} 
 	\mathrm{Tr} \left\{\hat{\Omega}[\mathcal{K}] 	(\theta,n )     \hat{g}  	 \right\}
 	 \mathrm{d}\Theta. 
 \end{split}
\label{2.21}
\end{equation}

 
\section{The generalized Moyal star product on the cylinder }\label{sec3}
Assume that the condition (\ref{2.9}) is fulfilled. Let $\hat{f}$ and $\hat{g}$ be operators in $L^2(S^1)$. We are on quest for the generalized Weyl symbol $W^{-1}[\mathcal{K}](\hat{f}\hat{g})$ of the product $\hat{f}\hat{g}$. 
From (\ref{2.19}), employing also the previous results on the Moyal star product on the cylinder \cite{16,16a} one has  
\begin{equation}
\begin{split}
W^{-1}[\mathcal{K}](\hat{f}\hat{g}) &= \left(\hat{K}^T \right)^{-1} W^{-1} [1](\hat{f}\hat{g}) = \left(\hat{K}^T \right)^{-1}\left( W^{-1} [1](\hat{f}) * W^{-1} [1](\hat{g}) \right)\\
&= 
 \left(\hat{K}^T \right)^{-1} \left[ \left( \hat{K}^T W^{-1}[\mathcal{K}](\hat{f}) \right) *  \left(\hat{K}^T W^{-1}[\mathcal{K}](\hat{g}) \right)  \right],
\end{split} 
\label{3.1}
\end{equation}
where the symbol $*$ stands for the usual Moyal star product on the cylinder 
\begin{equation}
\begin{split}
&\left( f*g \right) (\Theta,n\hslash) = \\
&=\frac{1}{4\pi^2} \sum _{n',n''=-\infty}^{\infty} \int_{-\pi}^{\pi}\int_{-\pi}^{\pi} f(\Theta',n'\hslash) 
\mathrm{Tr} \left\{ \hat{\Omega}[1](\Theta,n) \hat{\Omega}[1](\Theta',n' )
\hat{\Omega}[1](\Theta'',n'' ) \right\}
\times
g(\Theta'',n''\hslash) \mathrm{d}\Theta' \mathrm{d}\Theta'' \\
&
=\frac{1}{4\pi^2} \sum _{n',n''=-\infty}^{\infty} \int_{-\pi}^{\pi}\int_{-\pi}^{\pi}
f(\Theta',n'\hslash) 
\Big[\exp{ \left\{2i [(n''-n)(\Theta'-\Theta) - (n'-n)(\Theta''-\Theta) ]\right\}}\\
& 
\times 
\Big(
1+\mathrm{sgn}\big(\cos{(\Theta''-\Theta)}\big)\mathrm{sgn}\big(\cos{(\Theta'-\Theta)}\big)
+\mathrm{sgn}\big(\cos{(\Theta'-\Theta)})\mathrm{sgn}(\cos{(\Theta''-\Theta')}\big)\\
&+\mathrm{sgn}\big(\cos{(\Theta''-\Theta')})\mathrm{sgn}(\cos{(\Theta''-\Theta)}\big)
\Big)
g(\Theta'',n''\hslash)\Big] \mathrm{d}\Theta' \mathrm{d}\Theta''
\\ &\stackrel{\mathrm{(formally)}}{=}
f(\Theta,L) \exp{\left\{ \frac{i\hslash}{2} 
 \overleftrightarrow{\mathcal{P}}\right\}} g(\Theta,L) \Big|_{L=n\hslash},
\end{split}\label{3.2}
\end{equation}
 where $
 \overleftrightarrow{\mathcal{P}}$ is the Poisson operator
\begin{equation}
 \overleftrightarrow{\mathcal{P}} = \frac{\overleftarrow{\partial}}{\partial \Theta}   \frac{\overrightarrow{\partial}}{\partial L} - \frac{\overleftarrow{\partial}}{\partial L}  \frac{\overrightarrow{\partial}}{\partial \Theta}. 
\label{3.3}
\end{equation}
One quickly finds that Eq. (\ref{3.1}) with $\mathcal{K}=1$ under identification $W^{-1}[1](\hat{f})=f$ and $W^{-1}[1](\hat{g})=g$ yields 
\begin{equation}
W^{-1}[1](\hat{f}\hat{g})=f*g.
\label{3.4} 
\end{equation}
Therefore, it seems natural to introduce the \textit{generalized Moyal star product for the kernel $ \mathcal{K}$} denoted by    
 $ \underset{[\mathcal{K}]}{*} $ and defined by
 \begin{equation}
 f\underset{[\mathcal{K}]}{*} g := \left( \hat{K}^T\right) ^{-1} \left[ \left( \hat{K}^T f\right) * \left( \hat{K}^T g\right) \right].
 \label{3.5}
 \end{equation} 
Then Eq. (\ref{3.1}) can be rewritten in the form analogous to (\ref{3.4})
\begin{equation}
W^{-1}[\mathcal{K}] (\hat{f}\hat{g}) = W^{-1}[\mathcal{K}](\hat{f}) 
 \underset{[\mathcal{K}]}{*}W^{-1}[\mathcal{K}](\hat{g}) .
\label{3.6}
\end{equation}
Finally, the \textit{generalized Moyal bracket for the kernel $\mathcal{K}$} is given as
\begin{equation}
\left\{ f,g\right\}^{[\mathcal{K}]}_{M} := \frac{1}{i\hslash} \left(  f  \underset{[\mathcal{K}]}{*} g - g  \underset{[\mathcal{K}]}{*} f  \right).\label{3.7} 
\end{equation}
From (\ref{3.6}) and (\ref{2.12b}) one gets the relation
\begin{equation}
W[\mathcal{K}] \left(f \underset{[\mathcal{K}]}{*} g\right)  = 
W[\mathcal{K}](f) \cdot W[\mathcal{K}](g).
\label{3.8}
\end{equation}
Employing (\ref{3.8}) we obtain from (\ref{3.7}) the result which is crucial in Dirac's approach to quantization 
\begin{equation}
W[\mathcal{K}] \left ( \left\{ f,g\right\}^{[\mathcal{K}]}_{M} \right) = \frac{1}{i\hslash} 
\Big[ 
W[\mathcal{K}] (f) ,W[\mathcal{K}] (g) 
\Big],
\label{3.9}
\end{equation}
where the bracket $[\cdot,\cdot ]$ stands for the commutator. 
Note that all the generalized Moyal star products are equivalent. 
 
 \section{Wigner function }\label{sec4}
 Let $\hat{\varrho}$ be a density operator of a particle on the circle. It satisfies the usual conditions
 \begin{subequations}
 \begin{eqnarray}
 \hat{\varrho}^+&=&  \hat{\varrho} \label{4.1a},\\
  \braket{\psi|  \hat{\varrho}|\psi } &\geq& 0 \quad \forall{\ket{\psi} \in L^2(S^1)} \label{4.1b},\\
  \mathrm{Tr} \left\{  \hat{\varrho} \right\} &=& 1.\label{4.1c}
 \end{eqnarray}
 \end{subequations}
For  any observable represented by the operator $\hat{f}$ the expectation value of this observable in the state $\hat{\varrho}$ is given by the well known formula
\begin{equation}
\braket{\hat{f}} = \mathrm{Tr} \left\{ \hat{f}\hat{\varrho } \right\}.
\label{4.2}
\end{equation}
 Assume that the operator $\hat{f}$ arises from some classical observable $f=f(\Theta,L)$ as the result of the generalized Weyl quantization rule (\ref{2.1}). So by (\ref{2.21}) the relation (\ref{4.2}) can be rewritten in the form
 \begin{equation}
\braket{\hat{f}} = \sum_{n=-\infty } ^{\infty} \int_{-\pi}^{\pi} f(\Theta, n\hslash) \frac{1}{2\pi} \mathrm{Tr} \left\{ \hat{\Omega} [\mathcal{K}](\Theta,n)   \hat{\varrho} \right\} \mathrm{d} \Theta.
  \label{4.3}
 \end{equation}
 Introducing the function
\begin{equation}
\varrho_W [\mathcal{K}] (\Theta, n\hslash)
:= 
\frac{1}{2\pi} \mathrm{Tr} \left\{ \hat{\Omega} [\mathcal{K}] (\Theta,n)   \hat{\varrho} \right\} 
\label{4.4}
\end{equation}
 which will be called the \textit{Wigner function for the state $\hat{\varrho}$ and the kernel $\mathcal{K}$} one writes (\ref{4.3}) as 
\begin{equation}
\braket{\hat{f}} = \sum_{n=-\infty}^{\infty} \int_{-\pi}^{\pi} f(\Theta,n\hslash) \varrho_W[\mathcal{K}](\Theta,n\hslash) \mathrm{d} \Theta.
\label{4.5}
\end{equation}
Eq. (\ref{4.5}) resembles very closely the fundamental formula from classical statistical mechanics defining the expectation value of the observable $f=f(\Theta,n\hslash)$. Therefore we can identify $\braket{\hat{f}} \equiv \braket{f(\Theta, n\hslash) } $.
 One easily finds that if the condition (\ref{2.6c}) is fulfilled then by (\ref{2.7b}) and (\ref{4.1a}) we have
 \begin{equation}
\varrho^*_W[\mathcal{K}] = \varrho_W[\mathcal{K}]
  \label{4.6}
 \end{equation}
i.e. $\varrho_W[\mathcal{K}]$ is a real function. 
Assume that (\ref{2.6a}) holds true. Performing summation over $n$ of both sides of (\ref{4.4}) and employing (\ref{2.7d}) one gets
\begin{equation}
\sum_{n=-\infty}^{\infty} \varrho_W[\mathcal{K}](\Theta,n\hslash) 
=
\mathrm{Tr} \left\{ 
\ket{\Theta} \bra{\Theta} \hat{\varrho} \right\} = \braket{\Theta|\hat{\varrho}|\Theta}
=:P(\Theta).
\label{4.7}
\end{equation} 
 The function $P(\Theta)$ given by (\ref{4.7}) is the \textit{probability distribution of the angle $\Theta$ in the state $\hat{\varrho}$}. 
Analogously, assuming (\ref{2.6b}), performing integration with respect to $\Theta$ of both sides of (\ref{4.4}) and, finally employing (\ref{2.7e}) one has 
\begin{equation}
\int_{-\pi} ^{\pi} \varrho_W[\mathcal{K}](\Theta,n\hslash) \mathrm{d} \Theta =
 \mathrm{Tr}  \left\{ \ket{n} \bra{n} \hat{\varrho} \right\}
= \braket{n|\hat{\varrho}|n} =: \mathcal{P}(n\hslash)  
\label{4.8}
 \end{equation} 
 i.e. the \textit{probability distribution of the angular momentum $L$ in the state $\hat{\varrho}$}.
 Note that any Wigner function should have the properties described by (\ref{4.6}), (\ref{4.7}) and (\ref{4.8}) \cite{3,7}.

 
 \section{Eigenvalue equations}\label{sec5}
 Given a Hermitian operator $\hat{f}=\hat{f}^{\dagger}$ in $L^2(S^1)$ the eigenvalue equation for $\hat{f}$ reads 
 \begin{equation} 
\hat{f} \ket{\psi} = \Lambda \ket{\psi} \implies \hat{f} \ket{\psi}\bra{\psi} = \Lambda \ket{\psi}\bra{\psi}, \, \Lambda \in \mathbb{R}, \quad
\Big[ \hat{f} ,\ket{\psi} \bra{\psi} \Big] =0
  \label{5.1}
 \end{equation}
 If we put $\braket{\psi|\psi}=1$, then the operator $\hat{\varrho} =\ket{\psi}\bra{\psi}$ is the density operator of the pure state represented by $\ket{\psi}$ and Eq. (\ref{5.1}) can be rewritten as 
 \begin{equation}
 \hat{f}\hat{\varrho} = \Lambda \hat{\varrho}, \quad \Lambda \in \mathbb{R}
 \label{5.2}
 \end{equation}
 with the extra constraint $[\hat{f},\hat{\varrho}]=0$.
 Assume that the condition (\ref{2.9}) is satisfied. 
 Then employing (\ref{3.1}) with (\ref{2.19}) and (\ref{4.4}) one easily gets 
 \begin{equation}
 \hat{K}^T W^{-1}[\mathcal{K}] (\hat{f}) * \hat{K}^{-1} \varrho_W[\mathcal{K}] = \Lambda  \hat{K}^{-1} \varrho_W[\mathcal{K}].\label{5.3}
\end{equation}
 In particular for $\mathcal{K}=1$   equation (\ref{5.3}) takes a simple form 
 \begin{equation}
 W^{-1}[1](\hat{f}) * \varrho_W[1] = \Lambda \varrho_W[1]. \label{5.4}
 \end{equation}
 
 
 \section{Symmetric ordering of operators}\label{sec6}
 
Wigner functions of a particle on the circle for $\mathcal{K}=1$ (the \textit{Weyl ordering}) have been considered in \cite{8,9,10,15,16,18}. In particular, recently \cite{18} the Wigner functions corresponding to the coherent states on the circle have been studied. 
However, as shown in the next section, if one intends to carry over the results on the Wigner functions in the cylindrical phase space $S^1\times \mathbb{R}^1$ to the case of number-phase Wigner functions in quantum optics it is more convenient to deal with the kernel which used in (\ref{2.1}) leads to the symmetric ordering of operators. As it has been shown in \cite{26,27,28,24,1} such a kernel reads 
\begin{equation}
\mathcal{K}(\sigma,\lambda) = \cos{\left( \frac{\sigma \lambda}{2} \right)} \equiv \mathcal{K}_S(\sigma,\lambda) \label{6.1}
\end{equation} 
and it fulfills all the conditions (\ref{2.6a}),(\ref{2.6b}) and (\ref{2.6c}).
Therefore the relations (\ref{2.7a}), (\ref{2.7b}), (\ref{2.7c}), (\ref{2.7d}) and (\ref{2.7e}) are also satisfied. Moreover, the respective  Wigner function $\varrho_W[\mathcal{K}_S]$ has the basic properties (\ref{4.6}), (\ref{4.7}) and (\ref{4.8}). Substituting (\ref{6.1}) into (\ref{2.3}) one quickly gets 
\begin{equation}
\begin{split}
\hat{\Omega} [\mathcal{K}_S ](\Theta,n) &=\frac{1}{2} \sum_{k=-\infty} ^ {\infty} \Big[ \exp{\left\{ -i(n-k)\Theta \right\} }\ket{n} \bra{k}  + \exp{\left\{ i(n-k)\Theta \right\} }\ket{k} \bra{n} \Big] \\
&= \pi \big[[ 
\ket{n} \braket{n|\Theta}  \bra{\Theta} + \ket{\Theta} \braket{\Theta| n} \bra{n} 
\big],
\end{split}\label{6.2}
\end{equation}
where
\begin{equation}
\ket{\Theta} = \frac{1}{\sqrt{2\pi}} \sum_{k=-\infty} ^{\infty}  \exp{ \left\{  -ik\Theta \right\} }\ket{k}
\label{6.3}
\end{equation}
is the normalized eigenvector of the angle operator $\hat{\Theta }$ and $\braket{n|\Theta}=\frac{1}{\sqrt{2\pi}}\exp{\left\{ -in\Theta \right\}}$ (for detailed analysis of the angle operator $\hat{\Theta}$ see for instance  \cite{29,1}).
Inserting (\ref{6.2}) into (\ref{4.4}) we have 
\begin{equation}
\begin{split}
\varrho_W[\mathcal{K}_S](\Theta,n\hslash) &= \frac{1}{2\pi} \mathrm{Re} \left[ \sum_{k=-\infty} ^ {\infty} \exp{\left\{ -i(n-k) \Theta\right\}} \braket{k|\hat{\varrho}|n} \right] \\
&= \mathrm{Re} \left[  \braket{\Theta| \hat{\varrho}|n} \braket{n|\Theta}  \right].
\end{split}
\label{6.4}
\end{equation}
Observe now that 
\begin{equation}
\cos{\frac{\sigma l}{2}}=0 \Leftrightarrow \sigma l =(2j+1)\pi,\quad j,l\in\mathbb{Z}.
\label{6.5}
\end{equation}
Hence, the condition (\ref{2.9}) is not fulfilled and, consequently, the right hand side of (\ref{2.19}) is, in general, not well defined for $\hat{f} = \hat{\varrho}$. Nevertheless, the function $\varrho_W[\mathcal{K}_S](\Theta,n\hslash)$ given by (\ref{6.4}) defines the density operator $\hat{\varrho}$ uniquely. Indeed, one can rewrite (\ref{6.4}) in the following form
\begin{equation}
\varrho_W[\mathcal{K}_S](\Theta,n\hslash) =\frac{1}{2\pi} \sum_{k=-\infty}^{\infty} \Big[ 
\cos{[(n-k)\Theta] } \mathrm{Re} \braket{k| \hat{\varrho}|n} + \sin{[(n-k)\Theta] } \mathrm{Im} \braket{k| \hat{\varrho}|n}
 \big].
\label{6.6}
\end{equation}
From (\ref{6.6}) we easily get
\begin{subequations}
\begin{eqnarray}
\mathrm{Re} \braket{k|\hat{\varrho}|n} &=& 2 \int_{-\pi}^{\pi} \varrho_W [\mathcal{K}_S ] (\Theta,n\hslash) \cos{[(n-k)\Theta]} \mathrm{d}\Theta \quad \mathrm{for} \, k\neq n,\label{6.7a}\\
\braket{n|\hat{\varrho}|n} &=& \int_{-\pi} ^{\pi} \varrho_W[\mathcal{K}_S](\Theta,n\hslash) \mathrm{d}\Theta, \label{6.7b}\\
\mathrm{Im} \braket{k|\hat{\varrho}|n} &=& 2 \int_{-\pi}^{\pi} \varrho_W [\mathcal{K}_S ] (\Theta,n\hslash) \sin{[(n-k)\Theta]} \mathrm{d}\Theta \label{6.7c}
\end{eqnarray}
\end{subequations}
(of course (\ref{6.7b}) follows also from (\ref{4.8})).
Multiplying both sides of (\ref{6.7c}) by $i=\sqrt{-1}$ and adding to (\ref{6.7a}) one finally has 
\begin{eqnarray}
\braket{k|\hat{\varrho}|n} &=& 2 \int_{-\pi}^{\pi} \varrho_W [\mathcal{K}_S ] (\Theta,n\hslash) \exp{[i(n-k)\Theta]} \mathrm{d}\Theta \quad \mathrm{for} \, k\neq n,\nonumber \\
\braket{n|\hat{\varrho}|n} &=& \int_{-\pi} ^{\pi} \varrho_W[\mathcal{K}_S](\Theta,n\hslash) \mathrm{d}\Theta. \label{6.8}
\end{eqnarray}
Concluding, the formulas (\ref{6.8}) give all matrix elements of $\hat{\varrho}$ in the angular momentum basis $\left\{   \ket{n} \right\}_{n=-\infty}^{\infty}$, \textit{ipso facto} the density operator $\hat{\varrho}$ itself.

 
 \section{Number-phase Wigner function}\label{sec7}
 
 The problem of defining the quantum phase of a harmonic oscillator or of a single-mode electromagnetic field has a long and involved history initiated by P. A. M. Dirac \cite{30} and F.~London \cite{31} in the years $1926$-$27$. There is not a place here to discuss all meanders of that history and we refer the reader to some of numerous works devoted to this question \cite{21,22,23,32,33,34,35,36,37,38,39,40,41,42,43,44,45,46,47,1}.
In the present paper we employ the results of our recent work \cite{1} where it is argued that a reasonable way to define the quantum phase consists in extending the Fock space of the harmonic oscillator or the single-mode electromagnetic field to the Hilbert space $L^2(S^1)$. This idea has been considered by several authors  \cite{39,40,41,42,43,46}. The construction given in Ref. \cite{1} can be stated as follows. Let $\mathcal{H} _F$ be the respective Fock space and denote the \textit{Fock basis of $\mathcal{H}_F$}  by 
 $\left\{ \ket{\underline{n}} \right\}_{n=0}^{\infty}$.
 (\textit{Remark:} The vectors of $\mathcal{H}_F$ will be marked by the additional under-bar i.e.  $\ket{\underline{n}}$, $\ket{\underline{\psi}}$, $\ket{\underline{\chi}}$,\dots   etc.)
 We embed $\mathcal{H}_F$ in the Hilbert space $L^2(S^1)$ by
 \begin{equation}
 \hat{J}: \mathcal{H}_F \ni \sum_{n=0}^{\infty} c_n \ket{\underline{n}} \mapsto 
 \sum_{n=0}^{\infty} c_n \ket{n} \in L^2(S^1), \quad c_n\in\mathbb{C}.
 \label{7.1}
 \end{equation}
 Define the projection $\hat{\Pi}$ of $L^2(S^1)$ onto $\mathcal{H}_F$ by
\begin{equation}
\hat{\Pi} : \left\{
\begin{array}{rl}
L^2(S^1) \ni \ket{n} \mapsto \ket{\underline{n}} \in \mathcal{H}_F,
& n=0,1,2,\dots\\
L^2(S^1) \ni \ket{n} \mapsto 0 \in \mathcal{H}_F, & n=-1,-2,\dots
\end{array} \right. \label{7.2}
\end{equation}
and put
\begin{subequations}
 \begin{eqnarray}
 \bra{\psi} \hat{\Pi} &:=& \left( \hat{\Pi} \ket{\psi} \right)^{\dagger}\label{7.3a} \\
  \bra{\underline{\chi}} \hat{J} &:=& \left( \hat{J} \ket{\underline{\chi}} \right)^{\dagger}. \label{7.3b}
 \end{eqnarray}
 \end{subequations}
Then for any classical observable being a function of the phase $\phi$, $f=f(\phi)$ one assigns the corresponding quantum observable in a state $\ket{\underline{\psi}}\in \mathcal{H}_F$ by quantizing the classical observable $f(-\Theta)$ on the circle in the state $\hat{J}\ket{\underline{\psi}} \in L^2(S^1)$ using the generalized Weyl quantization rule (\ref{2.1}). The expectation value of the quantum observable corresponding to $f=f(\phi)$ in the state $\ket{\underline{\psi}}$ is equal to the expectation value of $f(-\Theta)$ calculated for $\hat{J}\ket{\underline{\psi}}\in L^2(S^1)$. As has been shown in \cite{1}, the procedure described above is equivalent to the approach developed by J.~H.~Shapiro and S.~R.~Shepard \cite{36}, and by P.~Bush, M.~Grabowski and P.~J.~Lahti \cite{38}, where the quantum phase is given as the \textit{positive operator valued (POV) measure} on $[-\pi,\pi)$
\begin{equation}
M_0 : \mathcal{B}\left( [-\pi,\pi) \right) \ni X \mapsto \frac{1}{2\pi} 
\sum_{j,k=0}^{\infty} \left( \int_X \exp{\{ i(j-k) \phi\}}\mathrm{d} \phi  \right) \ket{\underline{j}} \bra{\underline{k}}
\label{7.4}
 \end{equation} 
 with $\mathcal{B}\left( [-\pi,\pi) \right) $  standing for the family of Borel sets on $[-\pi,\pi)$. 
 This POV measure is a \textit{compression} of the spectral measure $E$ i.e. $M_0(X)=\hat{\Pi} E(X) \hat{\Pi}$
 \begin{equation}
E: \mathcal{B}\left( [-\pi,\pi) \right) \ni X \mapsto \frac{1}{2\pi} 
\sum_{j,k=-\infty}^{\infty} \left( \int_X \exp{\{ i(j-k) \phi\}}\mathrm{d} \phi  \right) \ket{{j}} \bra{{k}}.
\label{7.5}
 \end{equation} 
 Thus the projection $\hat{\Pi}$ defined by (\ref{7.2}) is just a Naimark projection \cite{38,48,49}.
 Moreover, one can also show \cite{1} that our procedure is equivalent to the Pegg-Barnett approach \cite{21,22,23} but without  introducing any, rather artificial, finite-dimensional Hilbert spaces. 
 Now we are at the position when the number-phase Wigner function can be defined.
 Let 
 \begin{equation}
\hat{ \underline{\varrho}} =
 \sum_{j,k=0}^{\infty} \varrho_{jk}  \ket{\underline{j}} \bra{\underline{k} }
\label{7.6}
 \end{equation}
 be a density operator on the Fock space $\mathcal{H}_F$.
 Then the operator 
 \begin{equation}
\hat{ {\varrho}} = \hat{J} \hat{\underline{\varrho}} \hat{J} =
 \sum_{j,k=0}^{\infty} \varrho_{jk}  \ket{{j}} \bra{{k} }
\label{7.7}
 \end{equation} 
is the \textit{density operator on $L^2(S^1)$ associated to $\hat{\varrho}$}.
 Now the idea is to define the number-phase Wigner function corresponding to the state $\hat{\underline{\varrho}}$ (\ref{7.6}) on $\mathcal{H}_F$ by the  Wigner function 
 $\varrho_W[\mathcal{K}](\Theta,n\hslash)$ corresponding to the state $\hat{\varrho}$ (\ref{7.7}) on $L^2(S^1)$. More precisely, the \textit{number-phase Wigner function for the state $\hat{\underline{\varrho}}$ and the kernel $\mathcal{K}$} is defined as 
\begin{equation}
\underline{\varrho}_W [\mathcal{K}](\phi,n) := \varrho_W[\mathcal{K}] (-\phi,n\hslash) = 
\frac{1}{2\pi} \mathrm{Tr} \left\{  \hat{\Omega}[\mathcal{K}] (-\phi,n) \hat{\varrho}  \right\}, \quad n\geq 0.
\label{7.8}
\end{equation}
 Since in $\mathcal{H}_F$, $n=0,1,2,\dots$, for the sake of  further consistency one must assume that the kernel $\mathcal{K}$ is such that 
 \begin{equation}
\forall_{\Theta\in[-\pi,\pi)}, \forall_{j,k \geq 0,\,\, n < 0}\quad 
 \braket{j| \hat{\Omega} [\mathcal{K}](\Theta,n) | k}=0.
 \label{7.9}
 \end{equation}
 From  (\ref{2.3}) with (\ref{2.2}) we easily infer that the condition (\ref{7.9}) is equivalent to the following condition imposed on $\mathcal{K}$
 \begin{equation}
 \forall_{j,k \geq 0,\,\, n < 0} \quad \int_{-\pi} ^{\pi} \mathcal{K}(\sigma, j-k) \exp{\left\{ i \sigma \left( \frac{j+k}{2}-n \right) \right\}  }\mathrm{d} \sigma =0.
 \label{7.10}
 \end{equation}
 One quickly finds that this condition is not fulfilled for $\mathcal{K}=1$ (the Weyl ordering) and it is fulfilled, for example, for $\mathcal{K} = \mathcal{K}_S$.
 Therefore,  we restrict ourselves to this letter case.  
 From (\ref{7.8}) under (\ref{6.4}), (\ref{7.6}) and (\ref{7.7}) one obtains 
 \begin{equation}
 \begin{split}
\underline{\varrho}_W [\mathcal{K}_S](\phi,n) &= \frac{1}{2\pi} \mathrm{Re}
\left[ 
\sum_{k=0}^{\infty} \exp{\{ i(n-k) \phi\}}
\braket{\underline{k} | \hat{\underline{\varrho}} | \underline{n}}
\right]\\&=\mathrm{Re} \left[
\braket{\Theta = -\phi | \hat{\varrho} | n} \braket{n|\Theta = -\phi}
\right], \quad n=0,1,2, \dots .
\end{split}
  \label{7.11}
 \end{equation}
 Let $f=f(\phi)$ be a classical observable relevant to the phase. Then employing (\ref{4.5}), 
(\ref{7.8}) and (\ref{7.11}) we find that the expectation value of the quantum observable 
corresponding to $f(\phi)$  in the state $\hat{\underline{\varrho}}$ 
given by (\ref{7.6}) reads
      \begin{equation}
 \braket{f(\phi)} = \frac{1}{2\pi} 
 \sum_{n,k=0}^{\infty} 
 \int_{-\pi}^{\pi} 
 f(\phi) \exp{\{ i(n-k) \phi \} } \varrho_{kn} 
 \mathrm{d} \phi. 
 \label{7.12}
 \end{equation}
In particular, if $\hat{\underline{\varrho}}$ is a pure state 
 \begin{equation}
 \hat{\underline{\varrho}} =\ket{\underline{\psi} } \bra {\underline{\psi} }, \quad \braket{\underline{\psi} | \underline{\psi}}=1,\label{7.13}
 \end{equation}
 the formula (\ref{7.12}) can be rewritten in the form 
      \begin{equation}
 \braket{f(\phi)} = \frac{1}{2\pi} 
 \sum_{n,k=0}^{\infty} 
 \int_{-\pi}^{\pi} 
 f(\phi) \exp{\{ i(n-k) \phi \} } 
 \braket{\underline{k} | \underline{\psi} } \braket{\underline{\psi} | \underline{n} }
 \mathrm{d} \phi. 
 \label{7.14}
 \end{equation} 
 This result agrees perfectly with the respective result calculated within the Pegg-Barnett formalism (see Eq. ($71$) in \cite{1}; note that in that equation in \cite{1} the factor $\frac{1}{2\pi}$ is erroneously missing).
 Define the vector $\ket{\underline{\phi}}$  in the rigged Hilbert space of $\mathcal{H}_F$ by
 \begin{equation}
 \ket{\underline{\phi}} := \frac{1}{\sqrt{2\pi}} \sum_{n=0}^{\infty} 
 \exp{\{in\phi\}} \ket{\underline{n}}.
 \label{7.15}
 \end{equation}
 One quickly finds the relations 
 \begin{equation}
 \int_{-\pi} ^{\pi}  \ket{\underline{\phi}} \mathrm{d} \phi \bra{\underline{\phi}} = \hat {1} 
 \label{7.16}
 \end{equation}
 and 
 \begin{equation}
 \braket{\phi | \phi'} =
  \frac{1}{2} \delta^{(S)} ( \phi -\phi') + \frac{1}{4\pi} - \frac{i}{4\pi}
   \cot{\frac{\phi - \phi'}{2}}
   \label{7.17}
 \end{equation}
 (see \cite{34}).
 From (\ref{7.16}) we infer that any $\ket{\underline{\psi}} \in \mathcal{H}_F$
 can be written in the form
 \begin{equation}
 \ket{\underline{\psi}} = \int_{-\pi} ^{\pi} \braket{\underline{\phi} | \underline{\psi}}  
 \ket{\underline{\phi}} \mathrm{d} \phi.
 \label{7.18}
 \end{equation}
 Then Eq. (\ref{7.14}) can be rewritten as 
\begin{equation} 
\braket{f(\phi)} = \int_{-\pi}^{\pi} f(\phi) \left| \braket{\underline{\phi} | \underline{\psi}}  \right|^2 \mathrm{d} \phi.
\end{equation}
Consequently, the function $\psi(\phi):= \braket{\underline{\phi} | \underline{\psi}}$ 
 can be considered as the \textit{wave function in the phase representation} for the state $\ket{\underline{\psi}}$  and $| \braket{\underline{\phi} | \underline{\psi}}|^2 = |\psi(\phi)|^2$ is the respective phase probability distribution.
 Thus, the vector $\ket{\underline{\phi}}$ defined by (\ref{7.15}) is the \textit{phase state vector} as it is assumed in quantum optics (see e.g. \cite{31,34,44,45,46,47}).
 Finally, in terms of  $\ket{{\underline{\phi}}}$ the formula (\ref{7.12}) reads
 \begin{equation}
 \braket{f(\phi)} = \int_{-\pi} ^{\pi} f(\phi) \braket{\underline{\phi} | \hat{\underline{\varrho}}| \underline{\phi}}\mathrm{d} \phi 
 \label{7.21}
 \end{equation}
 and (\ref{7.11}) takes the form analogous to (\ref{6.4})
 \begin{equation}
 \underline{\varrho}_W [\mathcal{K}_S](\phi,n)= \mathrm{Re}
  \left[  
  \braket{ \underline{\phi} 
  | 
  \hat{\underline{\varrho}}
    | \underline{n}  }  
    \braket{\underline{n} | \underline{\phi}} 
 \right], \quad n=0,1,2,\dots
 \label{7.22}
 \end{equation}
 The marginal distributions calculated for the  number-phase Wigner function (\ref{7.22}) give the phase probability distribution 
 \begin{equation}
 \begin{split}
& \sum_{n=0} ^{\infty} \underline{\varrho}_W [\mathcal{K}_S ] (\phi,n) = 
\braket{\underline{\phi} | \hat{\underline{\varrho}}  | \underline{\phi}} \\
&= \frac{1}{2\pi} \left\{  1+ 2\mathrm{Re} \left[ \sum_{k,n=0}^{\infty} 
\exp{\{ i(n-k) \phi \}} \varrho_{kn}
  \right]   \right\} =: P(\phi)
 \label{7.23}
 \end{split}
 \end{equation}
 (compare with Eq. $(6.86)$ from \cite{47} ) and the photon number probability distribution 
 \begin{equation}
 \begin{split}
 \int_{-\pi}^{\pi } \underline{\varrho}_W [\mathcal{K}_S](\phi,n) \mathrm{d} \phi 
 = 
 \braket{\underline{n} | \hat{\underline{\varrho}}| \underline{n} }
 =:\mathcal{P}(n).
 \label{7.24}
 \end{split}
 \end{equation}
  
 
   \section{ Some examples of the number-phase Wigner functions }\label{sec8}
\subsection{The Fock state $\ket{\underline{N}}$} 
Inserting $\hat{\underline{\varrho}}=\ket{\underline{N}}\bra{\underline{N}}$ into (\ref{7.22}) and employing (\ref{7.15}) one gets 
\begin{equation}
\underline{\varrho}_W [\mathcal{K}_S](\phi,n) = \left| \braket{\underline{\phi}|\underline{n}} \right|^2 \delta_{Nn} =\frac{1}{2\pi} \delta_{Nn}
\label{8.1}
\end{equation}
and this is just the Wigner function found by J. A. Vaccaro \cite{3}.
Then, the phase probability distribution (\ref{7.23}) is now uniform
\begin{equation}
P(\phi) =\frac{1}{2\pi}
\label{8.2}
\end{equation}
and the photon number probability distribution (\ref{7.24}) is simply
\begin{equation}
\mathcal{P}(n) =\delta_{Nn}
\label{8.3}
\end{equation}
as should be for the Fock state $\ket{\underline{N}}$.
Note that $\underline{\varrho}_W [\mathcal{K}_S](\phi,n)= P(\phi)\mathcal{P}(n)$.

\subsection{Coherent states} 
\begin{figure}
\begin{center}
\includegraphics[ ]{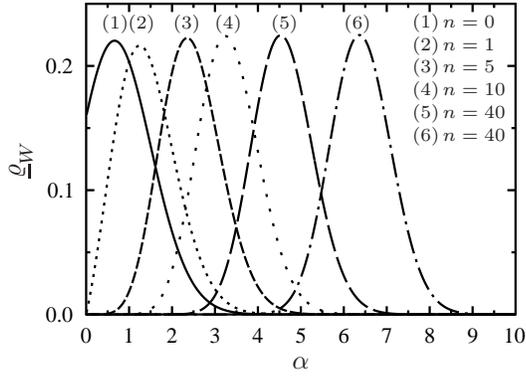}
\caption{\label{max}The maximum of $\underline{\varrho}_W$ ($\phi=0$) as a function of $\alpha=|\alpha|$ for some chosen $n$ varying from $0$ to $40$. }
\end{center}
\end{figure}

\begin{figure}
\begin{center}
\includegraphics[ ]{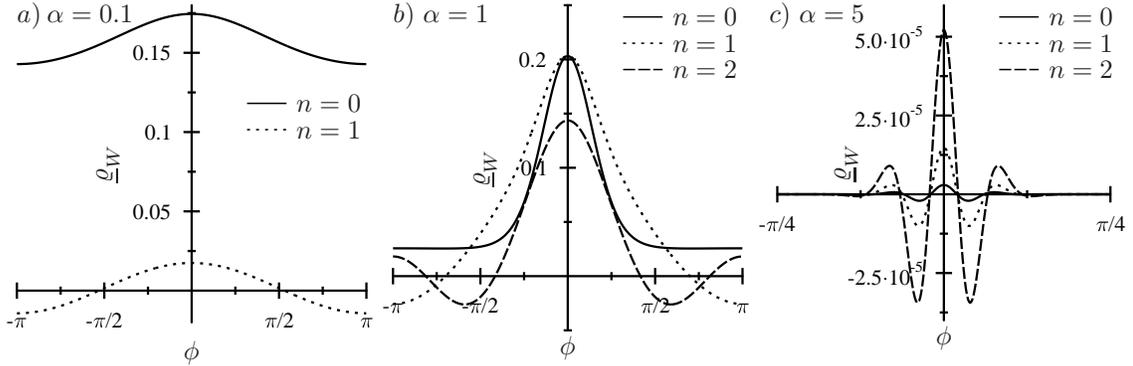}
\caption{\label{fig1} Exemplary plots of coherent state Wigner functions  $a)$ $\alpha=0.1$, $n=0$ and $n=1$, $b)$ $\alpha=1$, $n=0$, $n=1$ and $n=2$, $c)$ $\alpha=5$, $n=0$, $n=1$ and $n=2$  }
\end{center}
\end{figure}

\begin{figure}
\begin{center}
\includegraphics[ ]{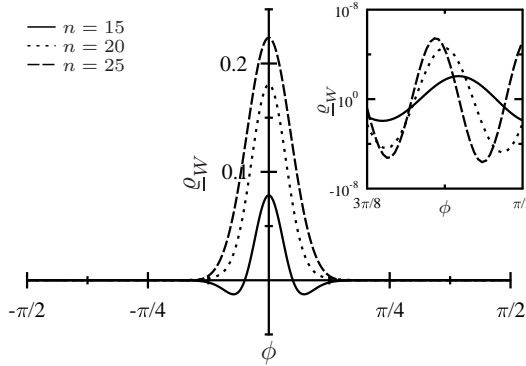}
\caption{\label{fig2} Coherent state Wigner  functions for $\alpha=5$, $n=15$, $n=20$ and $n=25$. Note the oscillations about zero after initial decay. }
\end{center}
\end{figure}

\begin{figure}
\begin{center}
\includegraphics[ ]{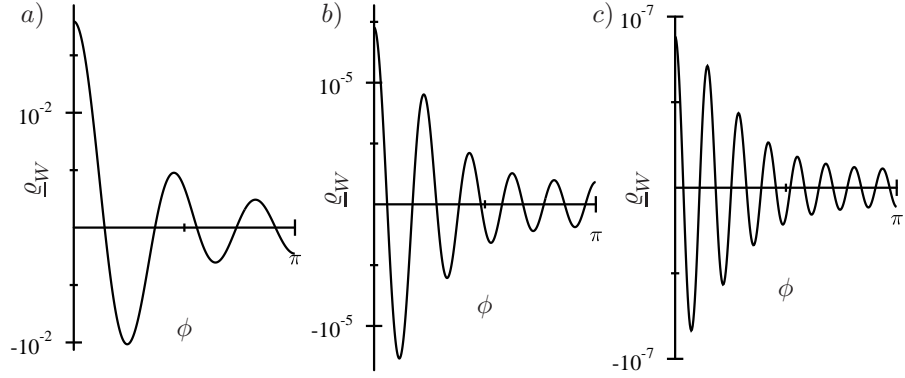}
\caption{\label{fig3} Coherent state Wigner  functions for $\alpha=1$,  $a)\,n=15$, $b)\, n=20$ and $c)\,n=25$. Note  the increase in the number of zeros with growing $n$. }
\end{center}
\end{figure}

Consider the coherent state
\begin{equation}
\ket{\underline{\alpha}} = \exp{\left\{ -\frac{1}{2} |\alpha|^2 \right\}}
\sum_{n=0}^{\infty} \frac{\alpha^n}{\sqrt{n!}} \ket{\underline{n}}, \quad \mathbb{C}\ni \alpha=|\alpha|e^{i\varphi}, \varphi\in\mathbb{R}.
\label{8.4}
\end{equation}
The corresponding density matrix is 
\begin{equation}
\hat{\underline{\varrho}} = \ket{\underline{\alpha}} \bra{\underline{\alpha}}.
\label{8.5}
\end{equation}
Substituting (\ref{8.5}) into (\ref{7.22}) we obtain
\begin{equation}
\begin{split}
\underline{\varrho}_W [\mathcal{K}_S] (\phi,n) &=
\frac{ |\alpha|^n\exp{ \left\{  -\frac{1}{2} |\alpha|^2 \right\} }}{\sqrt{2\pi n!} }
\mathrm{Re} \left\{  
\braket{\underline{\phi} | \underline{\alpha}} \exp{\{ in(\phi-\varphi)\}}
    \right\}\\
&= \frac{ |\alpha|^{2n}\exp{ \left\{  - |\alpha|^2 \right\} }}{2\pi\sqrt{ n!} }
\sum_{l=-n}^{\infty} \frac{|\alpha|^l}{\sqrt{(l+n)!}} \cos{\{ l(\phi-\varphi)\} }\\
&= \frac{ |\alpha|^n\exp{ \left\{  -  |\alpha|^2 \right\} }}{2\pi\sqrt{ n!} }
\Bigg[ 
\cos{\{ n(\phi-\varphi)\}}
\sum_{k=0}^{\infty} \frac{|\alpha|^k}{\sqrt{k!}} \cos{\{ k(\phi-\varphi)\} }
\\
&+
\sin{\{ n(\phi-\varphi)\}}
\sum_{k=0}^{\infty} \frac{|\alpha|^k}{\sqrt{k!}} \sin{\{ k(\phi-\varphi)\} }
\Bigg].
\end{split}
\label{8.6}
\end{equation}
It is clear from (\ref{8.6}) that if we put $\varphi=0$ (as we do in all the plots) the coherent state Wigner functions assumes its maximum for $\phi=0$. In the Figure \ref{max}  we plot the dependence of this maximum on $\alpha=|\alpha|$ for several values of $n$ from $0$ up to $40$.  We notice that as $n$ grows the peak position shifts towards higher values of $\alpha$.  Moreover, slight change in peak hight is only observed for small values of $n$. Figure \ref{fig1} shows the coherent state number-phase Wigner functions for $a)\, \alpha=0.1$, $n=0$ and $n=1$,
$b)\, \alpha=1$, $n=0$, $n=1$ and $n=3$ and
$c)\, \alpha=5$, $n=0$, $n=1$ and $n=3$.
We notice that for $n=0$ and small values of $\alpha$ considered Wigner functions are nonnegative. 
Figure \ref{fig2} shows coherent state Wigner functions for $\alpha=5$ and higher values of $n$, the sub-figure depicts the oscillatory character of the function.  
Finally, to show the increase of zeros with growing $n$, we plot the coherent state Wigner functions for $\alpha=1$ in the Figure \ref{fig3} $a)\,n=15$, $b)\,n=20$ and $c)\,n=25$.

\subsection{Squeezed stats}
\begin{figure}
\begin{center}
\includegraphics[ ]{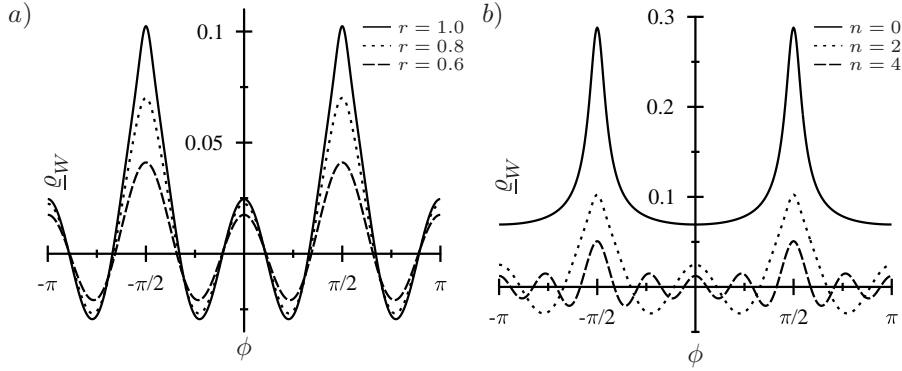}
\caption{\label{sv} Squeezed vacuum Wigner functions $a)$ for $n=2$ and $r=1$, $r=0.8$ and $r=0.6$}, $b)$ $r=1$, $n=0$, $n=2$ and $n=4$.
\end{center}
\end{figure}

The amplitude of the Fock state $\ket{\underline{n}}$ for the squeezed state  $\ket{ \underline{\alpha , \zeta}}$ is given as \cite{47,50}
\begin{equation}
\begin{split}
\braket{\underline{n}| \underline{\alpha , \zeta}} = \frac{1}{\sqrt{n! \mathrm{cosh} r}}
\left[ \frac{1}{2} e^{i\theta} \mathrm{tgh} r  \right] ^{n/2} 
\exp{\left[ - \frac{1}{2} \left\{ |\alpha|^2 + (\alpha^*)^2 e^{i\theta} \mathrm{tgh}r \right\} \right]}
H_n\left(
\frac{\alpha + \alpha^* e^{i\theta} \mathrm{tgh} r}{\sqrt{2 e^{i\theta}  \mathrm{tgh} r}}
  \right)
\end{split}
\end{equation}
where $H_n$ stands for Hermite polynomial of degree $n$,  
$\mathbb{C}\ni \alpha=|\alpha|e^{i\varphi}, \varphi\in\mathbb{R}$
is the coherent amplitude and   $\mathbb{C}\ni \zeta=re^{i\theta}, \theta,r\in\mathbb{R}$ is the squeeze parameter.
Equation (\ref{7.22}) gives following, rather complicated, expression for the Wigner function
\begin{equation}
\begin{split}
\underline{\varrho}_W[\mathcal{K}_S](\phi,n) &= 
\frac{1}{2\pi \mathrm{cosh} r} \mathrm{Re} 
\Bigg[ 
\frac{1}{\sqrt{n!}}
\exp{ \left\{  -|\alpha|^2 [1+ \cos (2 \varphi - \theta)  \mathrm{tgh}r ] \right\}}
\left( \frac{1}{2} e^{-i\theta} \mathrm{tgh} r   \right)^{n/2} \\
&\times H_n\left(
\frac{\alpha^* + \alpha e^{-i\theta} \mathrm{tgh} r}{\sqrt{2 e^{-i\theta}  \mathrm{tgh} r}}
  \right)
  \sum_{k=0}^{\infty} e^{i(n-k)\phi} \frac{1}{\sqrt{k!}} 
  \left( \frac{1}{2} e^{i\theta} \mathrm{tgh} r   \right)^{k/2}
  H_k\left(
\frac{\alpha + \alpha^* e^{-i\theta} \mathrm{tgh} r}{\sqrt{2 e^{i\theta}  \mathrm{tgh} r}}
  \right)
\Bigg].
\end{split}\label{sq1}
\end{equation}
In the case of squeezed vacuum ($\alpha=0$) formula (\ref{sq1}) for   $2n$ assumes the following form
\begin{equation}
\underline{\varrho}_W[\mathcal{K}_S](\phi,2n) = 
\frac{1}{2\pi \mathrm{cosh}r} \frac{\sqrt{(2n)!}}{n!}  \left(\frac{-\mathrm{tgh}r}{2}\right)^n
\sum_{l=0}^{\infty}
 \frac{\sqrt{(2l)!}}{l!}  \left(\frac{-\mathrm{tgh}r}{2}\right)^l
\cos{ \{ (n-l)(2\phi - \theta)\}}
\end{equation}
whereas for   $2n+1$ it reads
\begin{equation}
\underline{\varrho}_W[\mathcal{K}_S](\phi,2n+1) = 0.
\end{equation}
Figure \ref{sv} $a)$  shows the squeezed vacuum  number-phase Wigner function for $n=2$ 
and the real squeeze parameter ($\zeta=r$), $r=1$, $r=0.8$ and $r=0.6$.
 Part $b)$ of Figure \ref{sv} depicts the unsqueezed ($r=0$) number-phase Wigner function for $n=0$, $n=2$ and $n=4$. 
We observe that for $n=0$ and small $r$ ($r \leq 1$) the Wigner function is nonnegative.

\subsection{Black body radiation} 
The density operator for a single mode of the black body radiation is given by
\begin{equation}
\hat{\underline{\varrho}} = \sum_{n=0}^{\infty}
\left( 
1-e^{ -\beta \hslash \omega}
\right) e^{ -n\beta \hslash \omega }
\ket{\underline{n}}\bra{\underline{n}},
\label{8.7}
\end{equation}
where $\beta = \frac{1}{kT} $ and $\omega$ stands for the frequency of the mode. Inserting (\ref{8.7}) into (\ref{7.22}) one gets
\begin{equation}
\underline{\varrho}_W [\mathcal{K}_S](\phi,n ) = \frac{1}{2\pi}
\left( 
1-e^{ -\beta \hslash \omega}
\right) e^{ -n\beta \hslash \omega }
\label{8.8}
\end{equation}
Then the phase and number probability distributions are
\begin{equation}
P(\phi) = \frac{1}{2\pi}
\label{8.9}
\end{equation}
and
\begin{equation}
\mathcal{P}(n) = \left( 
1-e^{ -\beta \hslash \omega}
\right) e^{ -n\beta \hslash \omega } 
=
\frac{1}{\braket{n}+1} 
 \left( \frac{\braket{n}} {\braket{n}+1} \right)^n
\label{8.10}
\end{equation}
respectively; where 
$\braket{n} = \frac{1} { \exp{(\beta \hslash \omega)}-1}$.
Observe that analogously as in the case of the Fock state one has in the present case 
$\underline{\varrho}_W [\mathcal{K}_S](\phi,n)= P(\phi)\mathcal{P}(n)$.

\subsection{The `Fock cat' state}  
\begin{figure}
\begin{center}
\includegraphics[ ]{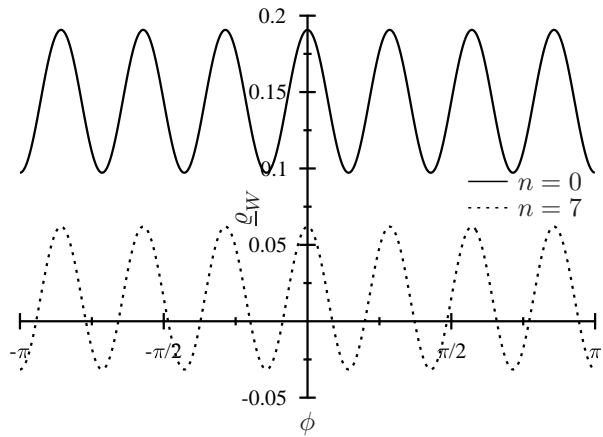}
\caption{\label{fc} The number-phase Wigner function   for the `Fock cat' state \newline $\ket{\underline{\psi}_{FC}} = \cos{(\pi/10)} \ket{\underline{0}} + \sin{(\pi/10)} \ket{\underline{7}}$.}
\end{center}
\end{figure}

We consider the `Fock cat' state in the form \cite{3}
\begin{equation}
\ket{\underline{\psi}_{FC}} = \cos{(\eta)} \ket{\underline{N}} +e^{i\varphi} \sin{(\eta)} \ket{\underline{N}'}.
\label{8.11}
\end{equation}
 Then
 \begin{equation}
\begin{split}
\hat{\underline{\varrho}} &=\ket{\underline{\psi}_{FC}}  \bra{\underline{\psi}_{FC}} = \cos^2(\eta) \ket{\underline{N}}\bra{\underline{N}} + 
\sin^2(\eta) \ket{\underline{N}'}\bra{\underline{N}'} \\
&+ \frac{1}{2} \sin{(2\eta)} \left\{ e^{i\varphi} \ket{\underline{N}'}\bra{\underline{N}}
 +e^{-i\varphi} \ket{\underline{N}}\bra{\underline{N}'}
    \right\}.
\end{split}
    \label{8.12}
 \end{equation}
Substituting (\ref{8.12}) into (\ref{7.22}) one gets
 \begin{equation}
\begin{split}
\underline{\varrho}_W [\mathcal{K}_s] (\phi,n) &= \frac{1}{2\pi} \Bigg\{ 
\cos^2(\eta)  \delta_{nN} + \sin^2(\eta)  \delta_{nN'}\\
&+ \frac{1}{2} \sin(2\eta)
 \cos{\left( (N-N') \phi +\varphi \right) (\delta_{nN} + \delta_{nN'})}
\bigg\}.
\end{split}
    \label{8.13}
 \end{equation}
In   Figure \ref{fc} we plot the Wigner function for the `Fock cat' state (\ref{8.11}) with $\eta=\pi/10$, $\varphi=0$, $N=0$ and $N'=7$. We quickly conclude that in contrast to the case  studied by J. A. Vaccaro \cite{3} where an interference ring is observed for  some $n$ between $n=N$ and $n=N'$, in our case the term with $n=N$ interferes with  $N'$ and \textit{vice versa} by the term $\frac{1}{2} \sin(2\eta)
 \cos{\left( (N-N') \phi +\varphi \right) (\delta_{nN} + \delta_{nN'})}$.

 
 \section{Concluding remarks}\label{sec9}
We have shown that using the concept of the enlarged Hilbert space for the Fock space $\mathcal{H}_F$ one can easily define the number-phase Wigner function in quantum optics.  
In our case the enlarged Hilbert space of $\mathcal{H}_F$ is the Hilbert space $L^2(S^1)$.
The advantage of such a choice is that  it eliminates the need of using any unphysical states with half integer values of the number of photons. Moreover, one can apply the elegant Weyl quantization formalism introduced for a particle on a circle $S^1$.
As a result we obtain a number-phase Wigner function which is real and its marginal distributions give the probability distributions of the phase and the number of photons, respectively. Further analysis and comparison of our results with the ones obtained by other authors will be given elsewhere.

{\bf Acknowledgments}
 M. P. and J. T. were partially supported by the CONACYT (Mexico) grant no. 103478. 
 

\end{document}